\documentclass[aps,prl,twocolumn,superscriptaddress,amsmath,amssymb]{revtex4-2}

\usepackage[utf8]{inputenc}
\usepackage{amsmath,amsthm}
\usepackage{graphicx}
\usepackage[colorlinks = true,
            linkcolor = blue,
            urlcolor  = blue,
            citecolor = blue,
            anchorcolor = blue]{hyperref}
\usepackage{braket}
\usepackage{xcolor}
\usepackage{easyReview}
\usepackage{bm}
\newcommand{\tr}[1]{\ensuremath{\operatorname{tr}\!{#1}}}

\newcommand{\bea}{\begin{eqnarray}}
\newcommand{\eea}{\end{eqnarray}}

\global\long\def\l{\la}

\newcommand{\vac}{{\rm vac}}

\def\RE{\mathop{\rm Re}} %
\def\IM{\mathop{\rm Im}} %

\global\long\def\ga{\gamma} \global\long\def\de{\delta}
\global\long\def\De{\Delta} \global\long\def\Ga{\Gamma}

\global\long\def\ell#1{\theta_{#1}}

\global\long\def\la{\lambda} 
\global\long\def\si{\sigma}

\global\long\def\eps{\epsilon}
\global\long\def\al{\alpha}
\global\long\def\be{\beta}
\global\long\def\ga{\gamma} \global\long\def\de{\delta}

\def\brho{\bm{\rho}}

\global\long\def\no{\nonumber}

\theoremstyle{remark}

\renewcommand{\thefigure}{\arabic{figure}}

\global\long\def\ell#1{\bar{\theta}_{#1}}

\begin{document}

\title{Bethe-ansatz diagonalization of steady state of boundary driven
  integrable spin chains}

\author{Vladislav Popkov} \affiliation{Faculty of Mathematics and
  Physics, University of Ljubljana, Jadranska 19, SI-1000 Ljubljana,
  Slovenia} \affiliation{Department of Physics, University of
  Wuppertal, Gaussstra\ss e 20, 42119 Wuppertal, Germany}
\affiliation{ These two authors contributed equally to this work}

\author{Xin Zhang} \affiliation{Beijing National Laboratory for
  Condensed Matter Physics, Institute of Physics, Chinese Academy of
  Sciences, Beijing 100190, China} \affiliation{ These two authors
  contributed equally to this work}

\author{Carlo Presilla} \affiliation{Dipartimento di Matematica,
  Universit\`a di Roma La Sapienza, Piazzale Aldo Moro 5, Roma 00185,
  Italy} \affiliation{Istituto Nazionale di Fisica Nucleare, Sezione
  di Roma 1, Roma 00185, Italy}

\author{Toma\v z Prosen} \affiliation{Faculty of Mathematics and
  Physics, University of Ljubljana, Jadranska 19, SI-1000 Ljubljana,
  Slovenia} \affiliation{Institute of Mathematics, Physics and
  Mechanics, Jadranska 19, SI-1000 Ljubljana, Slovenia}

\begin{abstract}
  We find that the density operator of non-equilibrium steady state (NESS) of XXZ 
  spin chain with strong ``sink and source"  boundary dissipation, can be described in terms of quasiparticles,  with renormalized---dissipatively
  dressed---dispersion relation.  The spectrum of the NESS is then
  fully accounted for by Bethe ansatz equations for an associated coherent
  system.   The dissipative dressing generates
  an extra singularity in the dispersion relation, which strongly
  modifies the NESS spectrum with respect to
  the spectrum of the corresponding coherent model.  
In particular,  this leads to a dissipation-assisted entropy reduction, due to the
  suppression -- in the NESS spectrum --  of plain wave-type Bethe states in
  favor of Bethe states localized at the boundaries.
\end{abstract}

\maketitle

Integrable many-body systems 
play an  important role in statistical mechanics.  For instance,  it was the celebrated
 Onsager solution of the 2D Ising model \cite{1944Onsager} that has proved, for the first time,  that phase transitions are related 
to  non-analyticity of the free energy functional,  giving rise to the theory of critical phenomena. 
 Solution of Korteweg-de Vries equation \cite{GGKM} has shown that a nonlinear system can possess an infinite number 
of local conservation laws,  which led to a concept of nonlinear asymptotic  Fourier Transform
\cite{1974Ablowitz} and to
  theory of stable nonlinear excitations -- solitons.  Integrable quantum many-body systems serve as reference 
points and as calibration tools for contemporary experiments in cold atoms \cite{2022Ketterle,2022KetterleHubbard}  and 
 in quantum circuits 
\cite{2024Abanin}.

One
practical definition of integrability for isolated quantum many-body
systems rests upon the possibility to find their spectra with
computational effort growing slower than exponentially in system size
$N$.  Indeed, finding specific solutions of the Bethe Ansatz equations for an
integrable Hamiltonian typically requires only ${\rm poly}(N)$
computation steps, while the dimension of the Hilbert space grows
exponentially in $N$.  This practical aspect of integrability is in full accordance with
the theoretical treatment of integrable quantum Hamiltonians,  which are centered around  finding 
 the Bethe Ansatz equations (BAE) for their spectra \cite{BaxterBook,HubbardBook,CaoBook}.

On the other hand,  the situation with the same quantum systems in nonequilibrium,  in particular, 
for systems exposed to dissipation,  is quite different.  The central object for nonequilibrium systems
is the nonequilibrium steady state (NESS) which is a (often unique)  fixed point of the time evolution.
While,  due to discovery of the Matrix product ansatz
method \cite{2015ProsenMPAreview,2022TriangularMPAZeno},  the $k$-point correlations in the NESS can be calculated in a
polynomial time,  finding the NESS spectrum or even specific eigenvalues
of nontrivial NESS has so far remained accessible by direct diagonalization only,
i.e.  still requiring exponentially (with systems size) growing effort. 

\begin{figure}
  \centering
 \includegraphics[width=0.95\columnwidth,clip]{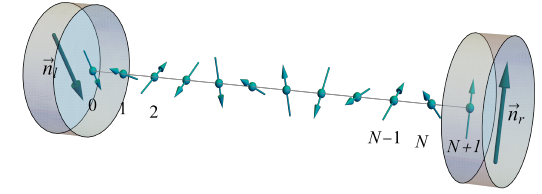}
  \caption{ Schematic picture of the dissipative setup.  The boundary spins are fixed by dissipation, while the 
internal spins follow an effective dynamics  consisting of
 fast coherent  dynamics  (\ref{XXZ}) and slow relaxation dynamics (\ref{ClassicalNESS}) towards the NESS.
For our ``sink and source" model (\ref{LME}) $\vec{n}_r=(0,0,1)$ and  $\vec{n}_l=-\vec{n}_r$.}
  \label{FigSchemaLetter}
\end{figure}

Our aim is to  present a method allowing to find the NESS spectrum of a famous quantum system -- 
XXZ spin chain -- 
attached to boundary dissipation of the ``sink and source" type,  via a Bethe Ansatz approach i.e.  in polynomial time. 
This demonstrates a principal possibiility to extend the above mentioned standard integrability paradigm to a certain family  of integrable  quantum systems in non-equilibrium.
In this communication we present a mathematically rigorous proof for our chosen ``sink and source" scenario
Fig.~\ref{FigSchemaLetter},   
while for other examples, where the validity of the 
same property is established numerically,  we direct a  reader to \cite{DressingPRA}.

One remarkable feature of our NESS spectrum solution that should be mentioned  beforehand,  is related to a 
concept of quasiparticles.  Quasiparticles are fundamental constituents of eigenstates of integrable coherent many-body systems: an energy $E_\al$ of any eigenstate $\ket{\al}$  is a sum of  individual  energies of all quasiparticles it contains: $E_\al = \sum_j \eps (u_{j\al})$
where $u_{j\al}$ is a set of quasiparticles rapidities and $\eps (u)$ is a dispersion relation.  

What we find is that the  NESS spectrum  $\{ \nu_\al \} = \{ \exp(-\tilde E_\al)\}$  can be written in 
the same form $\tilde E_\al = \sum_j \tilde \eps (u_{j\al})$,  where $u_{j\al}$ is a set of admissible rapidities   for
some related integrable coherent model,  and $\tilde \eps (u)$  is a new effective dispersion relation characterizing  ``coherent" quasiparticles ``renormalized" by the dissipation.  Such a ``survival" of  the intrinsically coherent 
objects -- quasiparticles --- under the action of dissipation, cannot be expected a-priori and is very surprising.

\textit{XXZ model with sink and source.---} Our aim is to find the spectrum of nonequilibrium steady state (NESS)
of a XXZ spin chain,  with boundary dissipation,  described by the 
 Lindblad Master equation for the  density matrix
\begin{align}
&\frac{\partial \brho(\Ga,t)}{\partial t} = - i [H,\brho] + \Ga \left( {\cal D }_{L_1}[\brho] +  {\cal D }_{L_2}[\brho]  \right)\label{LME}\\ 
&H =\sum_{n=0}^{N} \left( \si_{n}^x  \si_{n+1}^x + \si_{n}^y  \si_{n+1}^y+ \De \si_{n}^z \si_{n+1}^z  \right),\label{eq:fullXXZ}\\
&L_1=\si_0^-,\quad L_2=\si_{N+1}^{+},\label{eq:L1L2}\\
&{\cal D }_{L}[\brho] = L \brho L^\dagger - \frac12 \left(  L^\dagger L \brho + \brho  L^\dagger L \right). \label{LindbladOperator}
\end{align}
With the choice (\ref{eq:L1L2}) the  Lindblad dissipators ${\cal D }_{L_1}$  and  ${\cal D }_{L_2}$  
make the edge spins (spins at sites $0$ and $N+1$, see Fig.~\ref{FigSchemaLetter})  relax  
 towards  targeted states $\ket{\downarrow}$,  $\ket{\uparrow}$, with typical relaxation time $\tau_{\rm boundary} =O(1/\Ga)$.  The Matrix Product Ansatz for NESS (but not for NESS spectrum!) is formulated in 
\cite{2011Prosen}. 
Here we shall consider the so-called quantum Zeno (QZ) regime,  where $\tau_{\rm boundary}$ is much smaller than  typical time needed for the bulk relaxation $\tau_{\rm boundary}\ll \tau_{\rm bulk}$. 
 Then,   fast relaxation of the edge spins constrains the reduced density matrix of (\ref{LME}) to an approximately factorized form \cite{2014Zanardi} $\brho(\Ga,t) = \rho_l \otimes \rho(\Ga,t) \otimes \rho_r + O(1/\Ga)$ for $t\gg 1/\Ga$,  where $\rho(\Ga,t) \approx {\rm tr}_{0,N+1} \brho(t)$  is the reduced density matrix of 
interior spins.
We are interested in computing the spectrum $\{\nu_\al \}$  of the (unique) Zeno NESS, 
\begin{align}
&\rho_{\rm NESS} = \lim_{\Ga \rightarrow \infty} \lim_{t \rightarrow \infty} \rho(\Ga,t) =
\sum_{\al} \nu_{\al} \ket{\al} \bra{\al}.
\end{align}
To this end,  we establish three crucial properties: 

(i) $\rho_{\rm NESS}$ commutes with the Hamiltonian
\begin{align}
&\left[ \rho_{\rm NESS},H_D\right] =0, \label{eq:Comm}\\
& H_{D} =\sum_{n=1}^{N-1}\left( \si_{n}^x  \si_{n+1}^x + \si_{n}^y  \si_{n+1}^y+ \De \si_{n}^z \si_{n+1}^z  \right)\no\\
&\qquad 
+\De( \si_N^z -\si_1^z),  \label{XXZ}
\end{align}
(ii) $\nu_{\al}$ are given by stationary solution of an auxiliary Markov process
\begin{align}
\Ga \,\frac{d\nu_\alpha}{dt} =
  \sum_{\be \neq \al} w_{\be \al} \nu_\be
  - \nu_\al \sum_{\be \neq \al} w_{\al \be},
  \quad \al=1,2,\dots 
  \label{ClassicalNESS}
\end{align} 
with rates
\begin{align}
  w_{\be \al} =
  |\bra{\al} \si_1^{-} \ket{\be}|^2 +  |\bra{\al} \si_N^{+}\ket{\be}|^2,
  \label{rates}
\end{align}
where $\ket{\al},\ket{\be}$ are eigenstates of (\ref{XXZ}).

(iii) $\{ \nu_\al\}$ satisfy the detailed
balance condition
\begin{align}
  \nu_\al w_{\al \be} = \nu_\be w_{\be \al} .
  \label{eq:DetailedBalance}
\end{align}
Let us comment on all the properties (i), (ii), (iii).

To obtain (i) we expand the time-independent  steady state solution $\brho(\Ga)$ of (\ref{LME}) in powers of $1/\Ga$,
$\brho(\Ga) =\rho_0 + \Ga^{-1} \rho_1 + \Ga^{-2} \rho_2+\ldots$  Substituting into (\ref{LME}) we obtain the recurrence $i [H, \rho_k]=({\cal D }_{L_1}+  {\cal D }_{L_2})[\rho_{k+1}]$,  for $k=0,1,\ldots$  For the consistency of the 
recurrence $[H, \rho_k]$ must have no overlap with the kernel of the operator $({\cal D }_{L_1}+  {\cal D }_{L_2})$,
equivalent to ${\rm tr}_{0,N+1} [H,\rho_{k}] =0$.  In the leading order $k=0$,  a substitution of $\rho_0 = \rho_l \otimes \rho_{\rm NESS} \otimes \rho_r$   into ${\rm tr}_{0,N+1} [H,\rho_{0}] =0$ yields    (\ref{eq:Comm}).  
The Hamiltonian (\ref{eq:Comm}) is called a dissipation-projected 
Hamiltonian \cite{2014Zanardi} and can alternatively be obtained via a Dyson expansion.  

To establish (ii) one 
writes down a Dyson expansion of the $ \brho(\Ga,t)$ in  (\ref{LME}) using 
$1/\Ga$ as a perturbation parameter.  This leads to an effective Lindblad  evolution of the internal spins,  see Fig.~\ref{FigSchemaLetter},  of the form 
\begin{align}
&\frac{\partial \rho(\Ga,t)}{\partial t} = - i [H_{D},\rho] + \frac{1}{\Ga} \left( {\cal D }_{\si_1^{-}}[\rho] +  {\cal D }_{\si_{N}^{+}}[\rho]  \right) + O(\Ga^{-2}),\label{LMEeff}
\end{align}
containing a dominant coherent part with the Hamiltonian (\ref{XXZ}) and perturbative (of order $1/\Ga$) 
Lindbladian dissipator.  The latter gives rise to the auxiliary Markov process   (\ref{ClassicalNESS}), (\ref{rates}).
The procedure is explained  in detail in  \cite{2014Zanardi,2018ZenoDynamics},  and \cite{DressingPRA}, so we shall omit it here.

Finally,   the detailed balance (iii) Eq (\ref{eq:DetailedBalance}) is a special property related to integrability of 
(\ref{XXZ})
 and it is proved aposteriori,  see \cite{SM}.

Our strategy is  to use the integrability of the Hamiltonian     (\ref{XXZ}) in order to calculate the rates (\ref{rates})
and then the NESS spectrum via  (\ref{eq:DetailedBalance}).  Due to 
(\ref{eq:Comm}), $\rho_{\rm NESS}$ and $H_D$ in (\ref{XXZ}) can be diagonalized simultaneously.  

The Hamiltonian  (\ref{XXZ})  is treated in its more general
version (for arbitrary longitudinal boundary fields) in the pioneering
paper of Sklyanin~\cite{1988Sklyanin}. 
Due to $U(1)$ invariance,  (\ref{XXZ}) can be
block-diagonalized within blocks of fixed total magnetization
$\sum_{n=1}^N \si_n^z$. The
eigenvalues of the block with magnetization $N-2M$ are given by
\begin{align}
  &E_{\al} =(N-1)\De+ \sum_{j=1}^M \eps (u_{j,\al}),
    \label{eq:quasiparticles1}\\
 & \eps(u) 
  = \frac{-2 \sin^2 \ga}  {\sinh(u+\frac{i\ga}{2})\sinh(u-\frac{i\ga}{2}) }.
  \label{eq:eps(u)Trig}
\end{align} 
where $u_{j,\al}$, $j=1,\dots,M$, are Bethe roots satisfying the set
of BAE
\begin{align}
  & \frac{ u_j^{[-3]} }{ u_j^{[+3]} }
    \left( \frac{ u_j^{[+1]} }{ u_j^{[-1]} } \right)^{2N+1}
    = \prod_{\substack{k=1\\ k\neq j}}^M \prod_{\sigma=\pm1}
    \frac{(u_j+\sigma u_k)^{[+2]}}{(u_j+\sigma u_k)^{[-2]}}, 
    \label{eq:BAE}
\end{align} 
in which we defined $u^{[k]} = \sinh( u + i k \ga/2 )$, where $\cos \ga=\De$.

Using the Algebraic Bethe Ansatz machinery and some previous results \cite{2007Kitanine}
we are able to calculate the  rates $w_{\be \al}$ from (\ref{rates}),(\ref{eq:DetailedBalance}) and hence the NESS spectrum $\{ \nu_\al\}$. 
Details of the proof are given in  \cite{SM}.  Usually,  finding correlation functions  for open integrable spin chains is a difficult task,  and a result (if it exists) is expressed in terms of multiple integrals, infinite products, etc..,
see e.g.  \cite{SlavnovBook,1995Jimbo,2008Kitanine,2007Kitanine}.  In our case,  we succeded to obtain an  elegant 
explicit formula for the NESS spectrum, namely: 
\begin{align}
\nu_\al &=A \exp(-\tilde E_\al), \no\\ 
\tilde E_\al & = \sum_{j}  \tilde \eps(u_{j,\al})  \label{eq:tildeE} \\
  \tilde \eps(u)
  &= \log \left|\frac{\sinh(u+\frac{3i\ga}{2}) \ \sinh(u-\frac{3i\ga}{2})}
    {\sinh(u+\frac{i\ga}{2}) \  \sinh(u-\frac{i\ga}{2})} \right|
       \label{eq:eps(u)DissTrig}\\
  &= \log |1-\eps(u)\De|,\qquad \De=\cos \ga,\no
\end{align}
where $A$ is a normalization constant,  and  $\tilde E_\al,  \tilde \eps(u)$ in (\ref{eq:tildeE}), (\ref{eq:eps(u)DissTrig})
 are obvious dissipative analogs of the $E_\al$,  $\eps(u)$ in
 (\ref{eq:quasiparticles1}), (\ref{eq:eps(u)Trig}).

To make the analogy even more obvious,  we compare two reduced density matrices: 
a Gibbs state for the dissipation projected Hamiltonian (\ref{XXZ}) and the  NESS,  governed by  (\ref{XXZ}):
\begin{align}
  \begin{split}
    &\rho_\mathrm{Gibbs} = \frac{1}{Z}
      \sum_{\al}  e^{-\be E_\al} \ket{\al}  \bra{\al},
    \\
    &E_\al = \sum_j \eps (u_{j\al}),
  \end{split}
  \label{eq:rhoGibbs}\\
 \begin{split}
 &\rho_{{\rm NESS}}  =
      \frac{1}{\ \tilde{Z}}
      \sum_{\al} e^{- \tilde{E}_\al}  \ket{\al}  \bra{\al},\\
 &  \tilde{E}_\al =  \sum_j \tilde {\eps}(u_{j\al}). \label{eq:dissDressing}
  \end{split}
\end{align}

Note that both the eigenstates $\ket{\al}$ and the Bethe rapidities $u_{j,\al}$ (solutions of  
Bethe Ansatz (\ref{eq:BAE})) 
in (\ref{eq:rhoGibbs}) and (\ref{eq:dissDressing}) are the same.

\begin{figure}[tbp]
  \centering \includegraphics[width=0.23\textwidth]{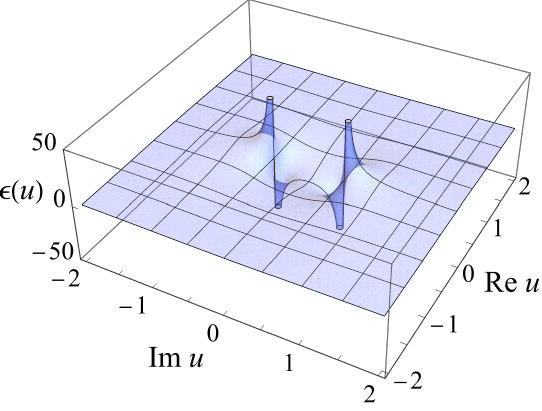}
  \includegraphics[width=0.23\textwidth]{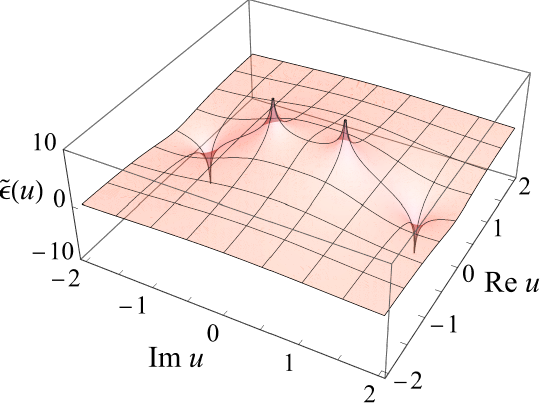}
  \caption{Surfaces $\eps(\RE u,\IM u)$ (left panel) and
    $\tilde \eps(\RE u,\IM u)$ (right panel)  for isotropic case $\De=1$ showing singularities at
    $u=\pm i/2$ and at $u=\pm i/2$, $u=\pm 3i/2$, respectively.  }
  \label{FigEsingularities}
\end{figure}

It is instructive to compare singularities of  the original dispersion $\eps(u) $ (\ref{eq:eps(u)Trig}) and of the $\tilde \eps(u) $
 (\ref{eq:eps(u)DissTrig}).
$\eps(u) $ has a singularity at $u_j = \pm i\ga/2$.  Its dissipative analog $\tilde \eps(u) $  (\ref{eq:eps(u)DissTrig})
inherits the singularity $u_j = \pm i\ga/2$ from $\eps(u)$ and ``dresses" $\eps(u)$ with an additional   singularity at
$u_j= \pm 3i\ga/2$, see also Fig.~\ref{FigEsingularities}.  For this reason we shall call  $\tilde \eps(u) $ a dissipatively 
dressed dispersion relation  and  $\tilde E_\al$ a dissipatively dressed energy.  In other words,  
$\eps(u_j)$ is an energy of a quasiparticle with rapidity $u_j$ and  $\tilde \eps(u_j) $ the energy of the same 
quasiparticle in the dissipative setup (Fig.~\ref{FigSchemaLetter}).

Eqs. (\ref{eq:tildeE}), (\ref{eq:eps(u)DissTrig}) are our main results.  We have shown 
that a spectrum of a NESS for the ``NESS-integrable" system (\ref{LME}) can be obtained by solving 
Bethe Ansatz equations for a related auxiliary coherent integrable  problem (\ref{XXZ}).   
In  \cite{DressingPRA} we argue that the effect of dissipative dressing of quasiparticles' dispersion
 is not restricted to the U(1)-symmetric XXZ model with sink and source (\ref{LME})
but it extends to other integrable spin chains,  with
 $U(1)$ symmetry broken at the boundary and in the bulk.   However,  for the latter models  a
rigorous  proof is currently lacking.

In the following   we discuss some physical consequences of the dissipative dressing (\ref{eq:eps(u)DissTrig}).

\textit{Dissipation assisted entropy reduction.---} 
Coupling of a quantum system to a dissipation typically leads to a mixed state and a volume-extensive 
entropy.   In setups with quenches,leading to steady bulk currents,  a volume-extensive entropy is also typical \cite{2007Aschbacher}, as
  well as in mesoscopic conductors\cite{2019JPN}. 
Here we show that subtle effect of dissipative dressing 
in the effective dispersion relation $\tilde{\eps}(u)$,  in the contrary,  can make the entropy subextensive in volume 
or even  vanishingly small,  see below.  

\begin{figure}
  \centering \includegraphics[width=0.23\textwidth]{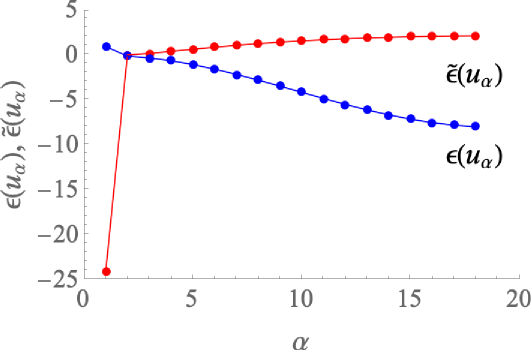}
  \includegraphics[width=0.23\textwidth]{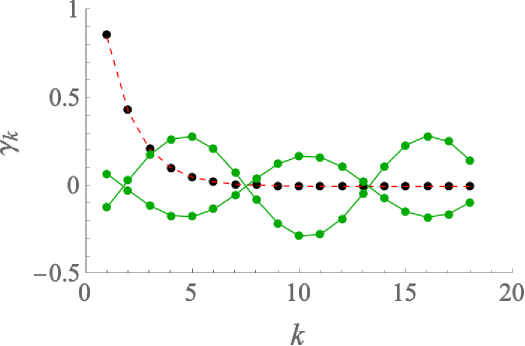}
  \caption{ Left panel: quasiparticle energies $\eps(u_\al)$ (blue
    joined points) and $\tilde \eps(u_\al)$ (red joined points) in the
    XXX model with $N=18$ spin, in the block with one magnon $M=1$.
    The state $\al=1$ is a localized Bethe state with
    $u_1\simeq 3i/2 +i e^{-24.5}$.  Right
    panel: coefficients $A_k$ of the normalized localized Bethe state
    $\ket{\al=1} = \sum_{k=1}^N A_k\ \si_k^{-} \ket{\uparrow\cdots\uparrow}$ (black empty
    circles). The dashed red line is the fit
    $A_k = 1.7 \times 2^{-k}$.  The green joined points are the
    coefficients $\RE A_k$ and $\IM A_k$ for the plain-wave like Bethe
    state with   $u_4\simeq 1.78139$. }
  \label{FigM1}
\end{figure}

To this end, we shall first discuss the isotropic case,  $\De=1$ in (\ref{XXZ}),  which 
is obtained from  the XXZ case by substituting $\ga \rightarrow \de$,
$u \rightarrow \de u $ and letting $\de \rightarrow 0$.  We then obtain  
BAE  (\ref{eq:BAE}) with $u^{[q]}=u + i q/2$,  while
$\eps(u), \tilde\eps(u)$ are given by
\begin{align}
\begin{split}
  &\eps (u) =  -\frac{2}{u^2+\frac{1}{4}},
    \\
& \tilde{\eps}(u) =
  \log \left| \frac{u^2 +\frac{9}{4}}{u^2 +\frac{1}{4}} \right|.
\end{split}
\label{eq:eps(u)}
\end{align} 
The surfaces $\eps(u), \tilde\eps(u)$ (\ref{eq:eps(u)}) for complex $u$ are
shown in Fig.~\ref{FigEsingularities}.  Let us consider  one-particle sector $M=1$, containing
$N$ Bethe eigenstates $\ket{\al} \equiv \psi(u_\al)$ parameterized by
the solutions $u_\al$, $\al=1,\ldots N$, of the BAE (\ref{eq:BAE}) with 
with $u^{[q]}=u + i q/2$.  
For any $N$ the set $\{u_\al\}$  consists of $N-1$ real
solutions, say $u_2, \ldots, u_N$ (corresponding to plane-wave type eigenstates) and one 
imaginary solution $u_1$  (corresponding to a boundary-localized eigenstates), see 
 Fig.~\ref{FigM1}.  The root $u_1$ lies exponentially close to the
singularity $u = 3 i/2$ of  $\tilde \eps(u)$ in (\ref{eq:eps(u)})  due to  dressing.  
Explicitly, from (\ref{eq:BAE}) we find
$u_1-3 i/2 \equiv \de = 3 i 2^{-2N-1} (1+ O(N\de))$,  see \cite{DressingPRA} for a proof.
The singularity of $\tilde\eps(u)$
 drastically decreases $\tilde\eps(u_1)$ with respect to ``coherent" energy $\eps(u_1)$,
the difference growing proportionally to the system size: $\tilde\eps(u_1) - \eps(u_1)
= -2(N+1)\log 2 + O(1)$.
  On the other hand, for real (plain wave type) solutions,
the dressed and original energies are comparable, see  Fig.~\ref{FigM1}.  As a result, 
the boundary localized Bethe eigenstate $\psi(u_1)$ enters the NESS (\ref{eq:dissDressing})
with an exponentially large relative weight with respect to the other eigenstates $\psi(u_2),  \ldots ,\psi(u_N)$ 
from the same one-quasiparticle sector $M=1$, see  Fig.~\ref{FigM1}.

\begin{figure}
  \centering \includegraphics[width=\columnwidth,clip]{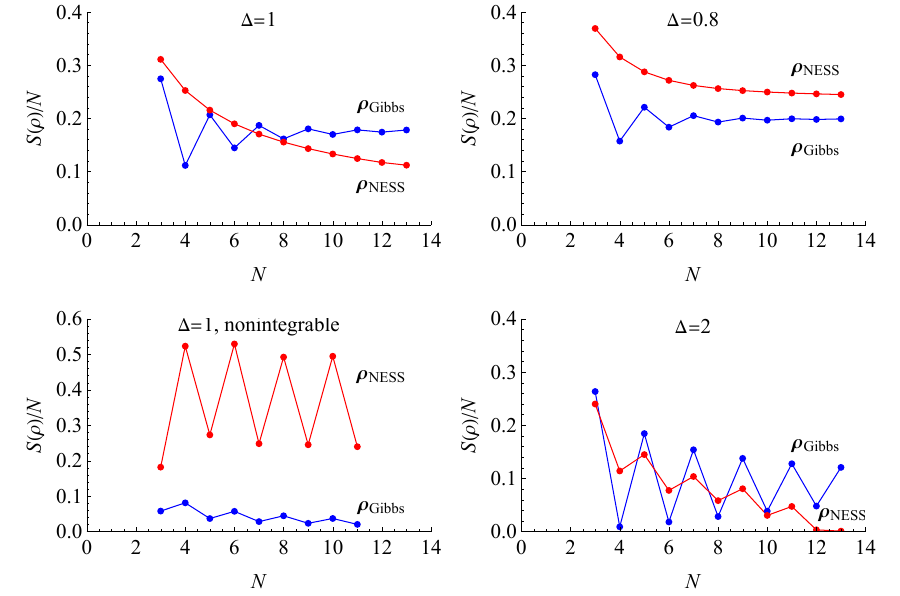}
  \caption{ Von Neumann entropy $S(\rho) = -\tr{(\rho \log \rho)}$ per
    spin versus the system size $N$ for $\rho=\rho_\mathrm{Gibbs}$
    (blue points) and $\rho=\rho_{\rm NESS}$ (red points),
    illustrating the dissipation assisted entropy reduction.  Left
    upper panel corresponds to the isotropic Heisenberg model
    $\Delta=1$, red dashed line is a fit given by $2/(3 N^{0.7}) $.
    Left bottom panel shows a nonintegrable case in which a staggered
    magnetic field $(-1)^j h\sigma^z_j$, with $h=1.5$, has been added.
    Right panels correspond to the anisotropic Heisenberg model in the
    easy plane $\Delta<1$ and easy axis $\Delta>1$ regime.  }
  \label{FigEntropy}
\end{figure}

The above mechanism predicting boundary-localized Bethe states to yield
dominant contribution to the NESS (\ref{eq:dissDressing}) can be qualitatively extended to higher
but fixed $M$.  E.g.  also in $M=2,3$ sectors there will be BAE roots exponentially
close to the singularity $u=\pm 3i/2$  of $\tilde \eps(u)$, corresponding to boundary-localized 
eigenstates of $H_D$.   
Understanding physics at fixed magnetization
density $M/N$ would require controlling thermodynamic Bethe ansatz in
the presence of boundary fields,  which is beyond our present scope.  
However,  we can see the net effect of dissipative dressing by looking at scaling of the von- Neumann entropy $S(\rho) = -\tr{(\rho \log \rho)}$
with system size $N$.  We observe the usual extensive volume law for Gibbs state (\ref{eq:rhoGibbs}),  at $\beta=1$, $S(\rho_{\rm Gibbs})= O(N)$
and  a clear sub-extensive scaling
$S(\rho_{\rm NESS}) \simeq N^{0.3}$ for the NESS (\ref{eq:dissDressing}),
see Left upper panel of  Fig.~\ref{FigEntropy}.  Keeping the dissipation term fixed,  and 
breaking the
integrability of (\ref{eq:fullXXZ}) (by adding a staggered magnetic field),  we ruin the subtle dressing mechanism,  rendering both $S(\rho_{\rm Gibbs})$,$S(\rho_{\rm NESS})$  volume-extensive, 
which is a conventionally expected outcome,   see 
left bottom panel of  Fig.~\ref{FigEntropy}.

\begin{figure}[t]
  \centering
  \includegraphics[width=\columnwidth]{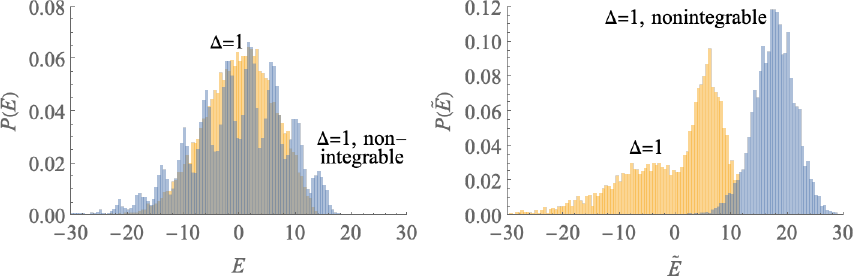}
  \caption{Distribution of the eigenvalues $E$ of
    $-\log \rho_\mathrm{Gibbs}$ (left panel) and $\tilde{E}$ of
    $-\log \rho_\mathrm{NESS}$ (right panel) for the XXX model with
    $N=13$.  In both panels, the yellow and blue histograms
    correspond, respectively, to the integrable and nonintegrable
    cases of Fig.~\ref{FigEntropy} with $\Delta=1$.  Distinct asymmetricity
    of $P(\tilde{E})$ (right Panel, yellow area) for the integrable case is due to the
    dissipative dressing effect. }
  \label{FigDistribEE}
\end{figure}

The difference of responses of an integrable and non-integrable spin chain
to the  boundary dissipation can be also well-seen on the distribution of eigenvalues 
 $\{E_\al\}$ and $\{\tilde{E}_\al\}$,  shown in  Fig.~\ref{FigDistribEE}.

The case $\De=1$ considered above lies at the boundary between the 
easy plane gapless regime $|\De|<1$ where correlations in XXZ model decay algebraically,
and the easy axis gapped regime with correlations
decaying exponentially with distance.  
The net effect of dressing (\ref{eq:eps(u)DissTrig}) in the two cases is also 
different.   Indeed, we observe  distinct behavior of the entropy of
NESS: a fast, perhaps
exponential decay of $S(\rho_{\rm NESS})/N$ in the gapped regime
$|\Delta|>1$, and saturation to a finite value (restoring  extensive volume law) $S(\rho_{\rm NESS})/N \sim {\rm const}$
in the gapless regime $|\Delta|<1$, see two right Panels of Fig. ~\ref{FigEntropy}.
For $\De=0$,  see Eq.(\ref{eq:eps(u)DissTrig}),  we have $\tilde \eps(u)=0$  
 leading to  $\rho_{\rm NESS}= I$ and $S(\rho_{\rm NESS})=N \log 2$. 
It is unclear at present if the 
 extensive volume law $S(\rho_{\rm NESS}) \sim O(N)$ sets in for  $|\De|<1$ or 
some critical value of $\De_{\rm crit}$ exists,  above which the  entropy
is sublinear.  Investigation of the  one-particle sector $M=1$, see \cite{DressingPRA}
shows that a boundary-localized eigenstate for large $N$, exponentially close to the singularity
$u=\pm 3i \ga/2$ in (\ref{eq:eps(u)DissTrig}) appears in the range $\frac12<|\De|\leq 1$
allowing to propose the lower bound $\De_{\rm crit}\geq \frac12$.

\textit{Discussion.---}We have developed an explicit Bethe ansatz
procedure for diagonalizing the steady state density operators of
boundary dissipatively driven integrable quantum spin chains in the
limit of large dissipation, alias the Zeno regime.  This becomes
possible due to a suprising phenomenon of ``dissipative dressing" of
quasiparticle energies in integrable coherent systems exposed to a
dissipation.
We find a general mechanism of entropy reduction due to
dissipation---pushing the steady state density matrix towards a pure
state---which is a consequence of additional singularities arising in
the quasiparticle dispersion relation due to dissipative dressing.

Our results should have applications in state engineering and
dissipative state preparation.  Moreover, we expect analogous emergent
integrability of the steady state in the discrete-time case of an
integrable Floquet XXX/XXZ/XYZ circuit, where boundary dissipation can
be conveniently implemented by the so-called reset
channel~\cite{2024Abanin}.

\begin{acknowledgements}
  V.P. and T.P. acknowledge support by ERC Advanced grant
  No.~101096208 -- QUEST, and Research Programme P1-0402 of Slovenian
  Research and Innovation Agency (ARIS). V.P. is also supported by
  Deutsche Forschungsgemeinschaft through DFG project
  KL645/20-2. X.Z. acknowledges financial support from the National
  Natural Science Foundation of China (No. 12204519).
\end{acknowledgements}



%

\clearpage

\onecolumngrid

\appendix

\setcounter{page}{1}
\setcounter{equation}{0}
\setcounter{figure}{0}

\renewcommand{\theequation}{\textsc{S}-\arabic{equation}}
\renewcommand{\thefigure}{\textsc{S}-\arabic{figure}}

\begin{center}
  {\bfseries {\em Supplemental Material  for ``Bethe-ansatz diagonalization of steady state of boundary driven
  integrable spin chains"}\\
  }
\end{center}

In the Supplemental Material we prove
 our main result, i.e. obtain the dissipatively-dressed dispersion relation for the $U(1)$ XXZ Hamiltonian
with diagonal boundary fields.

\section{S-I.  Derivation of Eq (\ref{eq:eps(u)DissTrig})}

We first recall some details of well-known Bethe Ansatz solution of an
open XXZ model with diagonal boundary fields
\begin{align}
  H=\sum_{n=1}^{N-1}[\sigma_{n}^x\sigma_{n+1}^x+\sigma_{n}^y\sigma_{n+1}^y+\cosh\eta\, \sigma_{n}^z\sigma_{n+1}^z]+\frac{\sinh\eta}{\tanh p}\, \sigma_1^z+\frac{\sinh\eta}{\tanh q}\sigma_N^z.\label{Ham1}
\end{align} 
Here, $\eta\equiv i\gamma$ is the anisotropic parameter.  The boundary fields we are interested in are 
$-\cosh\eta \sigma_1^z$ and $\cosh\eta \sigma_N^z$ (see Eq. (\ref{XXZ})), which corresponds to a substitution
$q=-p=\eta$ in (\ref{Ham1}) .  We shall perform the calculus for arbitrary $p,q$ and make the substitution at the very end. 

The $R$-matrix and $K$-matrices for the model are
\begin{align}
  R(u)&=\left(
        \begin{array}{cccc}
          \sinh (u+\eta) & 0 & 0 & 0 \\
          0 & \sinh u & \sinh \eta & 0 \\
          0 & \sinh \eta & \sinh u & 0 \\
          0 & 0 & 0 & \sinh (u+\eta) \\
        \end{array}
        \right),\\
  K^-(u)&=\left(
          \begin{array}{cc}
            \sinh (p+u-\frac{\eta}{2}) & 0 \\
            0 & \sinh (p-u+\frac{\eta}{2}) \\
          \end{array}
          \right),\\
  K^+(u)&=\left(
          \begin{array}{cc}
            \sinh (q+u+\frac{\eta}{2}) & 0 \\
            0 & \sinh (q-u-\frac{\eta}{2}) \\
          \end{array}
          \right).
\end{align}
Let us introduce the one-row monodromy matrices
\begin{align}
  \begin{aligned}
    T_0(u)&=R_{0,N}(u-\tfrac{\eta}{2})\ldots R_{0,1}(u-\tfrac{\eta}{2}),\\
    \hat T_0(u)&=R_{1,0}(u-\tfrac{\eta}{2})\ldots R_{N,0}(u-\tfrac{\eta}{2}),
  \end{aligned}
\end{align}
and the double-row monodromy matrix
\begin{align}
  \mathcal U_0^-(u)=T_0(u)K_0^-(u)\hat T_0(u)=\left(\begin{array}{cc}
                                                      A_-(u) & B_-(u) \\
                                                      C_-(u) & D_-(u) \\
                                                    \end{array}\right).
\end{align} 
The transfer matrix is given by
\begin{align}
  t(u)={\rm tr}_0\{K_0^+(u)\mathcal U^-_0(u)\}.
\end{align}
The Hamiltonian (\ref{Ham1}) rewritten in terms of the transfer matrix
reads
\begin{align}
  H=\sinh\eta\left.\frac{\partial \ln t(u)}{\partial u}\right|_{u=\frac{\eta}{2}}-N\cosh\eta-\tanh\eta\sinh\eta.
\end{align}
The eigenvalue of the quantum transfer matrix can be given by the
following $T-Q$ relation
\begin{align}
  \Lambda(u)=&\frac{\sinh(2u+\eta)}{\sinh(2u)}\prod_{s=p,q}\sinh(u+s-\tfrac{\eta}{2})\sinh^{2N}(u+\tfrac{\eta}{2})\prod_{j=1}^m\frac{\sinh(u-\l_j-\eta)\sinh(u+\l_j-\eta)}{\sinh(u-\l_j)\sinh(u+\l_j)}\no\\
             &+\frac{\sinh(2u-\eta)}{\sinh(2u)}\prod_{s=p,q}\sinh(u-s+\tfrac{\eta}{2})\sinh^{2N}(u-\tfrac{\eta}{2})\prod_{j=1}^m\frac{\sinh(u+\l_j+\eta)\sinh(u-\l_j+\eta)}{\sinh(u+\l_j)\sinh(u-\l_j)}.\label{TQ}
\end{align}
The Bethe roots $\{\l_1,\ldots,\l_m\}$ in (\ref{TQ}) satisfy the
following BAE
\begin{align}
  \left[\frac{\sinh(\l_j+\frac{\eta}{2})}{\sinh(\l_j-\frac{\eta}{2})}\right]^{2N}\prod_{s=p,q}\frac{\sinh(\l_j+s-\frac{\eta}{2})}{\sinh(\l_j-s+\frac{\eta}{2})}\prod_{k\neq j}^m\frac{\sinh(\l_j-\l_k-\eta)\sinh(\l_j+\l_k-\eta)}{\sinh(\l_j+\l_k+\eta)\sinh(\l_j-\l_k+\eta)}=1,\qquad j=1,,\ldots,m.\label{BAE}
\end{align}
Multiplying the left and right sides of BAE for all $j$, we get
\begin{align}
  \prod_{j=1}^m\left[\frac{\sinh(\l_j+\frac{\eta}{2})}{\sinh(\l_j-\frac{\eta}{2})}\right]^N\frac{\sinh^{\frac12}(\l_j+q-\tfrac{\eta}{2})\sinh^{\frac12}(\l_j+p-\tfrac{\eta}{2})}{\sinh^{\frac12}(\l_j-q+\tfrac{\eta}{2})\sinh^{\frac12}(\l_j-p+\tfrac{\eta}{2})}=\pm \prod_{1\leq j< k\leq m}\frac{\sinh(\l_j+\l_k+\eta)}{\sinh(\l_j+\l_k-\eta)}.\label{Product}
\end{align}
The energy of the system in terms of the Bethe roots reads
\begin{align}
  E=\sum_{j=1}^m\frac{2\sinh^2\eta}{\sinh (\l_j+\frac{\eta}{2}) \sinh (\l_j -\frac{\eta }{2})}+\sinh\eta(\coth p + \coth q)+(N-1)\cosh\eta.
\end{align} 
\subsection{Expressing local operators $\sigma_1^{\pm}$ and
  $\sigma_N^{\pm} $ in terms of monodromy matrix elements}
The trigonometric $R$-matrix possesses the following properties:
\begin{align}
  \begin{aligned}
    &R_{1,2}(0)=\sinh\eta P_{1,2},\\
    & R_{1,2}(u)R_{2,1}(-u)=\sinh(\eta+u)\sinh(\eta-u)\times \mathbb{I},\\
    &R_{1,2}(u)=-\sigma_1^yR^{t_1}_{1,2}(-u-\eta)\sigma_1^y.
  \end{aligned}
\end{align}
It is easy to check that
\begin{align}
  &{\rm tr}_0\{\sigma_0^-K_0^+(u)\mathcal U_0^-(u)\}=\sinh(q+u+\tfrac{\eta}{2})B_-(u),\\
  &{\rm tr}_0\{\sigma_0^+K_0^+(u)\mathcal U_0^-(u)\}=\sinh(q-u-\tfrac{\eta}{2})C_-(u).
\end{align}
We have
\begin{align}
  &{\rm tr}_0\{\sigma_0^-K_0^+(-\tfrac{\eta}{2})\mathcal U_0^-(-\tfrac{\eta}{2})\}\no\\
  &={\rm tr}_0\{[\sigma_0^-K_0^+(-\tfrac{\eta}{2})R_{0,N}(-\eta)\ldots R_{0,1}(-\eta)]^{t_0}[K_0^-(-\tfrac{\eta}{2})R_{1,0}(-\eta)\ldots R_{N,0}(-\eta)]^{t_0}\}\no\\
  &={\rm tr}_0\{[R_{0,N}(-\eta)\ldots R_{0,1}(-\eta)]^{t_0}[\sigma_0^-K_0^+(-\tfrac{\eta}{2})]^{t_0}[R_{1,0}(-\eta)\ldots R_{N,0}(-\eta)]^{t_0}[K_0^-(-\tfrac{\eta}{2})]^{t_0}\}\no\\
  &={\rm tr}_0\{R^{t_0}_{0,1}(-\eta)\ldots R^{t_0}_{0,N}(-\eta)[\sigma_0^-K_0^+(-\tfrac{\eta}{2})]^{t_0}R^{t_0}_{N,0}(-\eta)\ldots R^{t_0}_{1,0}(-\eta)[K_0^-(-\tfrac{\eta}{2})]^{t_0}\}\no\\
  &={\rm tr}_0\{R_{0,1}(0)\ldots R_{0,N}(0)[\sigma_0^y\sigma_0^-K_0^+(-\tfrac{\eta}{2})\sigma_0^y]^{t_0}R_{N,0}(0)\ldots R_{1,0}(0)[\sigma_0^yK_0^-(-\tfrac{\eta}{2})\sigma_0^y]^{t_0}\}\no\\
  &=\sinh^{2N}\!\!\eta[\sigma_N^y\sigma_N^-K_N^+(-\tfrac{\eta}{2})\sigma_N^y]^{t_N} {\rm tr}_0\{[\sigma_0^yK_0^-(-\tfrac{\eta}{2})\sigma_0^y]^{t_0}\}\no\\
  &=-2 \cosh \eta \sinh p \sinh q \sinh^{2N}\!\! \eta\,\sigma_N^-.\label{sigma;N-}
\end{align}
Analogously, we get
\begin{align}
  &{\rm tr}_0\{\sigma_0^+K_0^+(-\tfrac{\eta}{2})\mathcal U_0^-(-\tfrac{\eta}{2})\}=-2 \cosh \eta  \sinh p \sinh q \sinh^{2N}\!\! \eta\,\sigma_N^+.\label{sigma;N+}
\end{align}
Then, we get the following relation between $\sigma_N^\pm$ and
$B_-(-\tfrac{\eta}{2})$, $C_-(-\tfrac{\eta}{2})$
\begin{align}
  \sigma_N^-=-\frac{B_-(-\tfrac{\eta}{2})}{2\cosh\eta \sinh p \sinh^{2N}\!\! \eta},\\
  \sigma_N^+=-\frac{C_-(-\tfrac{\eta}{2})}{2\cosh\eta \sinh p \sinh^{2N}\!\! \eta}.
\end{align}
One can rewrite the transfer matrix in another way
\begin{align}
  t(u)&={\rm tr}_0\{[K_0^+(u)T_0(u)]^{t_0}[K_0^-(u)\hat T_0(u)]^{t_0}\}\no\\
      &={\rm tr}_0\{T_0^{t_0}(u)[K_0^+(u)]^{t_0}\hat T_0^{t_0}(u) [K_0^{-}(u)]^{t_0}\}\no\\
      &={\rm tr}_0\{[K_0^-(u)]^{t_0}[\mathcal{U}^+_0(u)]^{t_0}\}
        \no\\
      &={\rm tr}_0\{K_0^-(u)\mathcal{U}^+_0(u)\},
\end{align}
where
\begin{align}
  [\mathcal{U}^+_0(u)]^{t_0}&=T_0^{t_0}(u)[K_0^+(u)]^{t_0}\hat T_0^{t_0}(u)\no\\
                            &=R_{0,1}^{t_0}(u-\tfrac{\eta}{2})\ldots R_{0,N}^{t_0}(u-\tfrac{\eta}{2})[K_0^+(u)]^{t_0}R_{N,0}^{t_0}(u-\tfrac{\eta}{2})\ldots R_{1,0}^{t_0}(u-\tfrac{\eta}{2})\no\\
                            &=\left(\begin{array}{cc}
                                      A_+(u) & C_+(u) \\
                                      B_+(u) & D_+(u) \\
                                    \end{array}\right).
\end{align}
Using the same technique, we obtain that
\begin{align}
  \sinh p\, B_+(\tfrac{\eta}{2})&={\rm tr}_0\{\sigma_0^-K_0^-(\tfrac{\eta}{2})\mathcal{U}^+_0(\tfrac{\eta}{2})\}\no\\
                                &={\rm tr}_0\{[\sigma_0^-K_0^-(\tfrac{\eta}{2})]^{t_0}[\mathcal{U}^+_0(\tfrac{\eta}{2})]^{t_0}\}\no\\
                                &={\rm tr}_0\{[\sigma_0^-K_0^-(\tfrac{\eta}{2})]^{t_0}T_0^{t_0}(\tfrac{\eta}{2})[K_0^+(\tfrac{\eta}{2})]^{t_0}\hat T_0^{t_0}(\tfrac{\eta}{2})\}\no\\
                                &={\rm tr}_0\{K_0^+(\tfrac{\eta}{2})T_0(\tfrac{\eta}{2})\sigma_0^-K_0^-(\tfrac{\eta}{2})\hat{T}_0(\tfrac{\eta}{2})\}\no\\
                                &={\rm tr}_0\{K_0^+(\tfrac{\eta}{2})R_{0,N}(0)\ldots R_{1,0}(0)\sigma_0^-K_0^-(\tfrac{\eta}{2})R_{1,0}(0)\ldots R_{N,0}(0)\}\no\\
                                &=\sinh^{2N}\!\!\eta\, \sigma_1^-K_1^-(\tfrac{\eta}{2}){\rm tr}_0\{K_0^+(\tfrac{\eta}{2})\}\no\\
                                &=2\cosh \eta \sinh q \sinh p \sinh^{2N}\!\!\eta \,\sigma_1^-,\\
  \sinh p\, C_+(\tfrac{\eta}{2})&=2\cosh \eta \sinh q \sinh p \sinh^{2N}\!\!\eta  \,\sigma_1^+.
\end{align}
To find the rates of the auxiliary Markov process, see  the main text,  we need to calculate the quantity
\begin{align}
  \frac{|\bra{\mathbf{u}}\sigma_1^-\ket{\mathbf{v}}|^2+\bra{\mathbf{u}}\sigma_N^+\ket{\mathbf{v}}|^2}{|\bra{\mathbf{v}}\sigma_1^-\ket{\mathbf{u}}|^2+\bra{\mathbf{v}}\sigma_N^+\ket{\mathbf{u}}|^2}
  &=\frac{|\bra{\mathbf{u}}\sigma_1^-\ket{\mathbf{v}}|^2}{|\bra{\mathbf{u}}\sigma_N^-\ket{\mathbf{v}}|^2}=\left|\frac{\sinh p}{\sinh q}\right|^2\frac{|\bra{\mathbf{u}}B_+(\frac{\eta}{2})\ket{\mathbf{v}}|^2}{|\bra{\mathbf{u}}B_-(-\frac{\eta}{2})\ket{\mathbf{v}}|^2},\label{ratio1}
\end{align} 
where
$$\mathbf{u}=\{u_1,\ldots,u_{n+1}\},\qquad\mathbf{v}=\{v_1,\ldots,v_{n}\}$$
are the solution of BAE (\ref{BAE}) and $\ket{\mathbf{u}},\ket{\mathbf{v}} $ are the corresponding Bethe state.

\subsection{Calculating the expression (\ref{ratio1}) }

In the following we will recall some result of
Ref.~\cite{2007Kitanine} to derive the ratio
\begin{align}
  \frac{\bra{\mathbf{u}}B_+(\frac{\eta}{2})\ket{\mathbf{v}}}{\bra{\mathbf{u}}B_-(-\frac{\eta}{2})\ket{\mathbf{v}}}.\label{ratio2}
\end{align}
appearing in (\ref{ratio1}). The Bethe state $\ket{\bf u}$ can be
constructed by either $\prod_{j=1}^{n+1}B_-(u_j)\ket\vac$ or
$\prod_{j=1}^{n+1}B_+(u_j)\ket\vac$ where $\ket\vac=\ket{\uparrow\cdots\uparrow}$. Inserting the expression of
$\ket{\mathbf{v}}$, $\ket{\mathbf{u}}$ into (\ref{ratio2}), we get
\begin{align}
  \frac{\bra{\mathbf{u}}B_+(\frac{\eta}{2})\ket{\mathbf{v}}}{\bra{\mathbf{u}}B_-(-\frac{\eta}{2})\ket{\mathbf{v}}}
  &=\frac{\bra\vac C_-(u_1)\cdots C_-(u_{n+1})B_+(\tfrac{\eta}{2})B_+(v_1)\ldots B_+(v_n)\ket\vac}{\bra\vac C_+(u_1)\cdots C_+(u_{n+1})B_-(-\tfrac{\eta}{2})B_-(v_1)\ldots B_-(v_n)\ket\vac}\no\\
  &\qquad \times 
    \frac{\bra\vac C_+(u_1)\cdots C_+(u_{n+1})}{\bra\vac C_-(u_1)\cdots C_-(u_{n+1})}\frac{B_-(v_1)\ldots B_-(v_n)\ket\vac}{B_+(v_1)\ldots B_+(v_n)\ket\vac}.\label{ratio3}
\end{align}
From Ref. \cite{2007Kitanine}, we know that
\begin{align}
  &\frac{\bra\vac C_+(u_1)\cdots C_+(u_{n+1})}{\bra\vac C_-(u_1)\cdots C_-(u_{n+1})}\frac{B_-(v_1)\ldots B_-(v_n)\ket\vac}{B_+(v_1)\ldots B_+(v_n)\ket\vac}\no\\
  &=\prod_{j=1}^{n+1}\frac{\sinh(\eta-2u_j)}{\sinh(\eta+2u_j)}\prod_{j=1}^{n}\frac{\sinh(\eta+2v_j)}{\sinh(\eta-2v_j)}G(\{\mathbf{v}\};q,p)G(\{\mathbf{u}\};p,q),\label{ratio3-1}
\end{align}
where
\begin{align}
  G(\{\l_1,\ldots,\l_m\};x,y)=\prod_{j=1}^m\frac{\sinh^N(\l_j-\frac{\eta}{2})}{\sinh^N(\l_j+\frac{\eta}{2})}\frac{\sinh(\l_j-x+\frac{\eta}{2})}{\sinh(\l_j+y-\frac{\eta}{2})}\prod_{1\leq r< s\leq m}\frac{\sinh(\l_r+\l_s+\eta)}{\sinh(\l_r+\l_s-\eta)}.
\end{align}

With the help of Theorem 4.1 and Corollary 4.1 in \cite{2007Kitanine},
we get
\begin{align}
  &\frac{\bra\vac C_-(u_1)\cdots C_-(u_{n+1})B_+(\tfrac{\eta}{2})B_+(v_1)\ldots B_+(v_n)\ket\vac}{\bra\vac C_+(u_1)\cdots C_+(u_{n+1})B_-(-\tfrac{\eta}{2})B_-(v_1)\ldots B_-(v_n)\ket\vac}\no\\
  &=\frac{\bra\vac C_-(u_1)\cdots C_-(u_{n+1})B_+(\tfrac{\eta}{2})B_+(v_1)\ldots B_+(v_n)\ket\vac}{\bra\vac C_+(\frac{\eta}{2})C_+(-v_1)\cdots C_+(-v_n)B_-(-u_1)\ldots B_-(-u_{n+1})\ket\vac}
    \no\\
  &=\frac{\mathcal{S}_{n+1}^{-,+}(\{\mathbf{u}\};\{\tfrac{\eta}{2},\mathbf v\})}{\mathcal{S}_{n+1}^{+,-}(\{-\mathbf{u}\};\{\tfrac{\eta}{2},-\mathbf v\})},\label{ratio4}
\end{align}
where
\begin{align}
  &\mathcal{S}_m^{-,+}(\{\l\};\{\mu\})=\prod_{j=1}^m\sinh^N(\l_j-\tfrac{\eta}{2})\sinh^N(\l_j+\tfrac{\eta}{2})\,\frac{\det\,\mathcal{J}(\{\l\};\{\mu\})}{\det\,\mathcal{V}(\{\l\};\{\mu\})},\\
  &\mathcal{S}_m^{+,-}(\{\mu\};\{\l\})=\prod_{j=1}^m\sinh^N(\l_j+\tfrac{\eta}{2})\sinh^N(\l_j-\tfrac{\eta}{2})\,\frac{\det\,\mathcal{J}(\{\l\};\{\mu\})}{\det\,\mathcal{V}(\{\l\};\{\mu\})},\\
  &\mathcal{V}_{j,k}(\{\l\};\{\mu\})=\frac{\sinh(2\l_j)\sinh(2\mu_k-\eta)}{\sinh(2\l_j-\eta)\sinh(\mu_k-\l_j)\sinh(\mu_k+\l_j)},\\
  &\mathcal{J}_{j,k}(\{\l\};\{\mu\})=\frac{\partial}{\partial \l_j}\Lambda(\mu_k,\{\l\}).
\end{align}
The identity
\begin{align}
  \Lambda(u,\{\l\})=\Lambda(u,\{-\l\})=\Lambda(-u,\{-\l\}),
\end{align}
implies that
\begin{align}
  \frac{\det\,\mathcal{J}(\{\mathbf{u}\};\{\tfrac{\eta}{2},\mathbf v\})}{\det\,\mathcal{J}(\{-\mathbf{u}\};\{\tfrac{\eta}{2},-\mathbf v\})}=(-1)^{n+1}.
\end{align}
For the matrix $\mathcal{V}$, we have
\begin{align}
  &\frac{\mathcal{V}_{j,1}(\{\mathbf{u}\};\{x,\mathbf v\})}{\mathcal{V}_{j,1}(\{-\mathbf{u}\};\{x,-\mathbf v\})}=\frac{\sinh(2u_j+\eta)}{\sinh(2u_j-\eta)},\\
  &\frac{\mathcal{V}_{j,k\neq 1}(\{\mathbf{u}\};\{x,\mathbf v\})}{\mathcal{V}_{j,k\neq 1}(\{-\mathbf{u}\};\{x,-\mathbf v\})}=-\frac{\sinh(2v_{k-1}-\eta)}{\sinh(2u_j-\eta)}\frac{\sinh(2u_j+\eta)}{\sinh(2v_{k-1}+\eta)}.
\end{align}
As a consequence, it can be verified that
\begin{align}
  \frac{\det\,\mathcal{V}(\{\mathbf{u}\};\{\tfrac{\eta}{2},\mathbf v\})}{\det\,\mathcal{V}(\{-\mathbf{u}\};\{\tfrac{\eta}{2},-\mathbf v\})}=(-1)^{n}\prod_{j=1}^{n+1}\frac{\sinh(2u_j+\eta)}{\sinh(2u_j-\eta)}\prod_{k=1}^n\frac{\sinh(2v_k-\eta)}{\sinh(2v_k+\eta)}.
\end{align}
Then, one can derive the following equation
\begin{align}
  \frac{\mathcal{S}_{n+1}^{-,+}(\{\mathbf{u}\};\{\tfrac{\eta}{2},\mathbf v\})}{\mathcal{S}_{n+1}^{+,-}(\{-\mathbf{u}\};\{\tfrac{\eta}{2},-\mathbf v\})}=-\prod_{j=1}^{n+1}\frac{\sinh(2u_j-\eta)}{\sinh(2u_j+\eta)}\prod_{k=1}^n\frac{\sinh(2v_k+\eta)}{\sinh(2v_k-\eta)}.\label{ratio5}
\end{align}
Substituting (\ref{ratio3-1}) and (\ref{ratio5}) into (\ref{ratio3}),
we finally obtain
\begin{align}
  \frac{\bra{\mathbf{u}}B_+(\frac{\eta}{2})\ket{\mathbf{v}}}{\bra{\mathbf{u}}B_-(-\frac{\eta}{2})\ket{\mathbf{v}}}&=G^{-1}(\{\mathbf{v}\};q,p)G^{-1}(\{\mathbf{u}\};p,q)\no\\
                                                                                                                  &=\prod_{j=1}^{n+1}\frac{\sinh^N(u_j+\frac{\eta}{2})}{\sinh^N(u_j-\frac{\eta}{2})}\prod_{j=1}^{n+1}\frac{\sinh(u_j+q-\frac{\eta}{2})}{\sinh(u_j-p+\frac{\eta}{2})}\prod_{1\leq r<s\leq m}\frac{\sinh(u_r+u_s-\eta)}{\sinh(u_r+u_s+\eta)}\no\\
                                                                                                                  &\qquad\times \prod_{k=1}^{n}\frac{\sinh^N(v_k+\frac{\eta}{2})}{\sinh^N(v_k-\frac{\eta}{2})}\frac{\sinh(v_k+p-\frac{\eta}{2})}{\sinh(v_k-q+\frac{\eta}{2})}\prod_{1\leq r< s\leq n}\frac{\sinh(v_r+v_s-\eta)}{\sinh(v_r+v_s+\eta)}\no\\
                                                                                                                  &=\pm\prod_{j=1}^{n+1}\frac{\sinh^{\frac12}(u_j+q-\frac{\eta}{2})}{\sinh^{\frac12}(u_j-p+\frac{\eta}{2})}\frac{\sinh^{\frac12}(u_j-q+\tfrac{\eta}{2})}{\sinh^{\frac12}(u_j+p-\tfrac{\eta}{2})} \prod_{k=1}^{n}\frac{\sinh^{\frac12}(v_k+p-\frac{\eta}{2})}{\sinh^{\frac12}(v_k-q+\frac{\eta}{2})}\frac{\sinh^{\frac12}(v_k-p+\tfrac{\eta}{2})}{\sinh^{\frac12}(v_k+q-\tfrac{\eta}{2})}.
\end{align}

When $q=-p=\eta$, the Bethe ansatz equations are
\begin{align}
  &\left[\frac{\sinh(\l_j+\frac{\eta}{2})}{\sinh(\l_j-\frac{\eta}{2})}\right]^{2N+1}\frac{\sinh(\l_j-\tfrac{3\eta}{2})}{\sinh(\l_j+\tfrac{3\eta}{2})}\prod_{k\neq j}^m\frac{\sinh(\l_j-\l_k-\eta)\sinh(\l_j+\l_k-\eta)}{\sinh(\l_j+\l_k+\eta)\sinh(\l_j-\l_k+\eta)}=1,\qquad j=1,\ldots,m,\label{BAE3}\\
  {\rm or}\,&\left[\frac{\sinh(\l_j+\frac{i\gamma}{2})}{\sinh(\l_j-\frac{i\gamma}{2})}\right]^{2N+1}\frac{\sinh(\l_j-\tfrac{3i\ga}{2})}{\sinh(\l_j+\tfrac{3i\gamma}{2})}\prod_{k\neq j}^m\frac{\sinh(\l_j-\l_k-i\ga)\sinh(\l_j+\l_k-i\gamma)}{\sinh(\l_j+\l_k+i\ga)\sinh(\l_j-\l_k+i\ga)}=1,\qquad j=1,\ldots,m.\label{BAE3-1}
\end{align}
Then,
\begin{align}
  &\frac{|\bra{\mathbf{u}}\sigma_1^-\ket{\mathbf{v}}|^2+\bra{\mathbf{u}}\sigma_N^+\ket{\mathbf{v}}|^2}{|\bra{\mathbf{v}}\sigma_1^-\ket{\mathbf{u}}|^2+\bra{\mathbf{v}}\sigma_N^+\ket{\mathbf{u}}|^2}\no\\
  &=\left|\prod_{j=1}^{n+1}\frac{\sinh(u_j+\frac{\eta}{2})}{\sinh(u_j+\frac{3\eta}{2})}\frac{\sinh(u_j-\tfrac{\eta}{2})}{\sinh(u_j-\tfrac{3\eta}{2})}\prod_{k=1}^{n}\frac{\sinh(v_k-\frac{3\eta}{2})}{\sinh(v_k-\frac{\eta}{2})}\frac{\sinh(v_k+\tfrac{3\eta}{2})}{\sinh(v_k+\tfrac{\eta}{2})}\right|\no\\
  &=\left|\prod_{j=1}^{n+1}\frac{\sinh(u_j+\frac{i\ga}{2})}{\sinh(u_j+\frac{3i\gamma}{2})}\frac{\sinh(u_j-\tfrac{i\ga}{2})}{\sinh(u_j-\tfrac{3i\ga}{2})}\prod_{k=1}^{n}\frac{\sinh(v_k-\frac{3i\ga}{2})}{\sinh(v_k-\frac{i\ga}{2})}\frac{\sinh(v_k+\tfrac{3i\ga}{2})}{\sinh(v_k+\tfrac{i\ga}{2})}\right|,\label{w;ab}
\end{align}
leading  to Eq.~(\ref{eq:eps(u)DissTrig}).

\section{S-II.   Kolmogorov property of the rates $w_{\alpha \beta}$.  Proof of Eq.(\ref{eq:DetailedBalance})}

With the help of (\ref{w;ab}), we can derive that
\begin{align}
  w_{\alpha\beta}=\mathcal{F}_\alpha/\mathcal{F}_\beta,
\end{align}
where $\mathcal{F}_\alpha$ depends only on the  Bethe
roots, corresponding to eigenstate  $\ket{\alpha}$.  Therefore, we conclude that the following Kolmogorov
relation holds,
\begin{align}
  w_{\alpha\beta_1}w_{\beta_1\gamma}w_{\gamma\beta_2}w_{\beta_2\alpha}=w_{\alpha\beta_2}w_{\beta_2\gamma}w_{\gamma\beta_1}w_{\beta_1\alpha}. \label{Kolmogorov;1}
\end{align}
Eq.~(\ref{eq:DetailedBalance}) is a direct consequence of  the Kolmogorov relation (\ref{Kolmogorov;1}).

\end{document}